\documentclass[doublecol]{epl2}

\newcommand{\kB}{k_{\rm B}}
\newcommand{\expec}[1]{{\rm E}\big(#1\big)}
\newcommand{\Dbra}[1]{\left\langle#1\right|}
\newcommand{\Dket}[1]{\left|#1\right\rangle}
\newcommand{\Qcommu}[2]{[#1,#2]}
\newcommand{\Cdissip}[2]{\mbox{$\bm{[\hspace{-0.28em}[}$}#1,#2\mbox{$\bm{]\hspace{-0.28em}]}$}}

\title{Stochastic process behind nonlinear thermodynamic quantum master equation}
\shorttitle{Stochastic process behind thermodynamic quantum master equation}

\author{Hans Christian \"Ottinger}

\institute{ETH Z\"urich, Department of Materials, Polymer Physics, HCI H 543,
CH-8093 Z\"urich, Switzerland}


\abstract{We propose a piecewise deterministic Markovian jump process in Hilbert space such that the covariance matrix of this stochastic process solves the thermodynamic quantum master equation. The proposed stochastic process is particularly simple because the normalization of the vectors in Hilbert space is preserved only on average. As a consequence of the nonlinearity of the thermodynamic master equation, the construction of stochastic trajectories involves the density matrix as a running ensemble average. We identify a principle of detailed balance and a fluctuation-dissipation relation for our Markovian jump process.}

\begin{document}
\bibliographystyle{eplbib}

\maketitle

\section{Introduction}
Dissipative quantum systems are often described in terms of quantum master equations, where linear master equations of the Lindblad form are most popular in the literature \cite{BreuerPetru,Weiss,Lindblad76}. However, it has been known for some 30 years that these linear master equations have a fundamental problem because they invoke an inappropriate ``quantum regression hypothesis'' \cite{Grabert82,GraberTalkner83,Talkner86}. For quantum systems in contact with a heat bath, this problem has been overcome by the projection-operator derivation of a nonlinear quantum master equation associated with a ``modified quantum regression hypothesis'' \cite{Grabert82}. For more general thermodynamic environments, a generalized nonlinear quantum master equation has been obtained by a most natural extension of the geometric formulation of nonequilibrium thermodynamics from classical to quantum systems \cite{hco199,hco200,hco201} inspired by Dirac's method of classical analogy \cite{Dirac}.

The present letter addresses a possible route to the numerical solution of nonlinear thermodynamic quantum master equations. Stochastic simulation techniques have been established as a particulary versatile and convenient tool for solving quantum master equations \cite{GardinerZoller,BreuerPetru}. The basic idea is the \emph{unraveling} of the master equation (see, for example, chap.~6 of \cite{BreuerPetru}). The time-dependent density matrix or statistical operator $\rho(t)$ solving the master equation is obtained as a second moment or expectation,
\begin{equation}\label{unravel}
    \rho(t) = \expec{\Dket{\psi(t)} \Dbra{\psi(t)}} ,
\end{equation}
where $\psi(t)$ is a suitably defined stochastic process in the underlying Hilbert space. Such a process consists of random quantum jumps and a deterministic Schr\"odinger-type evolution modified by a friction term.

We begin with a brief summary of the general form of nonlinear quantum master equations based on thermodynamic principles. After introducing a general class of piecewise deterministic Markovian jump processes in the underlying Hilbert space, we identify the parameters required to reproduce a thermodynamic master equation and we discuss the form and significance of the jump operator in detail. In the conclusions, we discuss the usefulness of the Markovian jump process both for stochastic simulations and for gaining fundamental insights into dissipative quantum systems.

\section{Thermodynamic quantum master equation}
Based on purely thermodynamic considerations and a generalization from classical to quantum systems inspired by a geometric formulation of nonequilibrium thermodynamics, the following master equation for the evolution of the density matrix $\rho$ on a suitable Hilbert space has been proposed to characterize a quantum subsystem in contact with an arbitrary classical nonequilibrium system acting as its environment \cite{hco199,hco200,hco201}:
\begin{eqnarray}
    \frac{d\rho}{dt} &=& \frac{i}{\hbar} \Qcommu{\rho}{H}
    - \frac{1}{\kB} \Cdissip{H_{\rm e}}{S_{\rm e}}^Q_x \, \Qcommu{Q}{\Qcommu{Q}{H}_\rho}
    \nonumber\\
    && \qquad\qquad - \Cdissip{H_{\rm e}}{H_{\rm e}}^Q_x \, \Qcommu{Q}{\Qcommu{Q}{\rho}} .
\label{GENERICme}
\end{eqnarray}
In this equation, $\hbar$ and $\kB$ are Planck's constant (divided by $2\pi$) and Boltzmann's constant, respectively. The first term describes the reversible contribution to the evolution generated by the Hamiltonian $H$ via the commutator $\Qcommu{\,}{\,}$. The remaining terms are of irreversible nature and result from a coupling of the quantum subsystem to its classical environment. They are expressed through double commutators involving the self-adjoint coupling operator $Q$ so that the normalization condition, ${\rm tr}\,\rho=1$, is automatically preserved in time. The subscript $\rho$ on an arbitrary operator $A$ indicates the modified operator
\begin{equation}\label{Atildef}
    A_\rho = \int_0^1  \rho^\lambda A \, \rho^{1-\lambda} \, d\lambda ,
\end{equation}
which is basically the product of $A$ and $\rho$, but with a compromise between placing $\rho$ to the left or the right of $A$. If $A$ is self-adjoint, this property is inherited by $A_\rho$. We further have the useful identity
\begin{equation}\label{Arhoident}
    \Qcommu{A_\rho}{\ln\rho} = \Qcommu{A}{\rho} ,
\end{equation}
which can be established by evaluating the matrix elements for the eigenstates of $\rho$ \cite{hco199}.

Whereas the type of the coupling to the environment is given by the observable $Q$, the strength of the coupling is expressed in terms of a dissipative bracket $\Cdissip{\,}{\,}^Q$ defined as a binary operation on the space of observables for the classical environment (in this letter, boldface bracket symbols are used to distinguish classical dissipative brackets from quantum commutators). If the equilibrium or nonequilibrium states of the environment are characterized by state variables $x$, classical observables are functions or functionals of $x$, and their evaluation at a particular point of the state space is indicated by a subscript $x$. The classical observables $H_{\rm e}$ and $S_{\rm e}$ in eq.~(\ref{GENERICme}) are the energy and the entropy of the environment, respectively. For arbitrary environmental variables $A_{\rm e}$ and $B_{\rm e}$, the dissipative bracket occurring in the master equation (\ref{GENERICme}) is of the general form \cite{hco99,hco100,hcobet}
\begin{equation}\label{Cdissipdef}
    \Cdissip{A_{\rm e}}{B_{\rm e}}^Q_x = \frac{\partial A_{\rm e}(x)}{\partial x}
    \, M_{\rm e}^Q(x) \, \frac{\partial B_{\rm e}(x)}{\partial x} ,
\end{equation}
where $M_{\rm e}^Q$ is a positive-semidefinite symmetric matrix. The size of the square matrix $M_{\rm e}$ is given by the number of state variables $x$.

As a straightforward generalization of eq.~(\ref{GENERICme}), several coupling operators $Q_j$ can be incorporated easily \cite{hco199,hco200,Grabert82}. The proposed construction of an underlying stochastic process can be generalized accordingly. The master equation (\ref{GENERICme}) describes the influence of a classical environment on a quantum subsystem. The equation describing the reverse influence of the quantum system on its environment has been given in \cite{hco199,hco200,hco201}.

\section{Stochastic process}
In order to achieve a representation (\ref{unravel}) for the solution of the quantum master equation (\ref{GENERICme}), we consider a piecewise deterministic Markovian jump process in the underlying Hilbert space. The construction described here is inspired by the work of Breuer and Petruccione \cite{BreuerPetru,BreuerPetru95a,BreuerPetru95b}, with the pioneering precursor papers \cite{Dalibardetal92,DumZollerRitsch92}, however, we do not insist on preserving the normalization of the wave function for each realization of a stochastic trajectory. For nonlinear master equations, it is perfectly natural to think in terms of mean field interactions and to preserve certain properties only on average.

\subsection{Definition of Markovian jump process}
A key step in the construction of the underlying stochastic process is the introduction of the jump operator, which we assume to be of the general form
\begin{equation}\label{jumpoperatorgen}
    \tilde{Q} = \alpha ( Q + \beta \, \Qcommu{Q}{H}_\rho \rho^{-1} ) ,
\end{equation}
where the real coefficients $\alpha$ and $\beta$ remain to be determined. Note that, contrary to the coupling operator $Q$, the jump operator $\tilde{Q}$ is not self-adjoint. Once we have defined the jump operator $\tilde{Q}$, we only need to specify the jump rate and the deterministic evolution between jumps to obtain a complete definition of our piecewise deterministic Markovian jump process in Hilbert space. We perform quantum jumps
\begin{equation}\label{jump}
    \psi \rightarrow \tilde{Q} \psi
    \quad \mbox{with rate } \gamma ,
\end{equation}
where the rate parameter $\gamma$ remains to be determined. Note that the wave function $\tilde{Q} \psi$ is not normalized. Finally, we propose the following modified Schr\"odinger equation for the deterministic evolution of the state $\psi$ between the random jumps,
\begin{equation}\label{deterministic}
    \frac{d \psi}{d t} = - \frac{i}{\hbar} H \psi + \Lambda \psi ,
\end{equation}
where, contrary to the Hamiltonian $H$, the linear operator $\Lambda$ need not be self-adjoint. Again, the normalization of the wave function $\psi$ is not preserved. Whereas the jumps in eq.~(\ref{jump}) correspond to the noise represented by the diffusion term in a classical Fokker-Planck equation, the modification of the Schr\"odinger equation by the linear operator $\Lambda$ in eq.~(\ref{deterministic}) is the counterpart of deterministic friction. The striking simplicity of the proposed stochastic process defined by eqs.~(\ref{jumpoperatorgen})--(\ref{deterministic}) should be noted.

\subsection{Master equation for second moments}
In order to identify the proper choice of the parameters $\alpha$, $\beta$, $\gamma$ and of the friction operator $\Lambda$ we write the evolution equation for the second moments which, in the spirit of the unraveling idea expressed in eq.~(\ref{unravel}), can be regarded as an evolution equation for the density matrix,
\begin{equation}\label{stochme1}
    \frac{d\rho}{dt} = \frac{i}{\hbar} \Qcommu{\rho}{H} + \Lambda \rho + \rho \Lambda^\dag
    + \gamma ( \tilde{Q} \rho \tilde{Q}^\dag - \rho ) ,
\end{equation}
where the jumps with rate $\gamma$ lead to a loss of the current configuration and the creation of new states distorted by means of the jump operator $\tilde{Q}$. After inserting the definition (\ref{jumpoperatorgen}) of $\tilde{Q}$, eq.~(\ref{stochme1}) can be rewritten as
\begin{equation}\label{stochme2}
    \frac{d\rho}{dt} = \frac{i}{\hbar} \Qcommu{\rho}{H}
    - \gamma\alpha^2\beta \, \Qcommu{Q}{\Qcommu{Q}{H}_\rho}
    - \frac{1}{2} \gamma\alpha^2 \, \Qcommu{Q}{\Qcommu{Q}{\rho}} ,
\end{equation}
provided that the linear operator $\Lambda$ is chosen as
\begin{equation}\label{Lambdaid}
    \Lambda = \frac{\gamma}{2} \Big [ 1 - \alpha^2 Q^2
    + \alpha^2\beta^2 \big(\Qcommu{Q}{H}_\rho \rho^{-1}\big)^2 \Big] .
\end{equation}
The occurrence of the desired dissipative terms in eq.~(\ref{stochme2}) justifies the selection of the jump operator $\tilde{Q}$ according to eq.~(\ref{jumpoperatorgen}). It is this jump operator that expresses the thermodynamic nature of the nonlinear quantum master equation on the level of the stochastic process.

Note that both the jump operator $\tilde{Q}$ in eq.~(\ref{jumpoperatorgen}) and the friction operator $\Lambda$ in eq.~(\ref{Lambdaid}) are expressed in terms of $Q$ and $\Qcommu{Q}{H}_\rho \rho^{-1}$. The implied relationship between $\tilde{Q}$ and $\Lambda$ may be regarded as a \emph{fluctuation-dissipation relation}.

We can now compare eqs.~(\ref{GENERICme}) and (\ref{stochme2}) to determine the remaining coefficients. We find
\begin{equation}\label{gamalphid}
    \gamma\alpha^2 = 2 \, \Cdissip{H_{\rm e}}{H_{\rm e}}^Q_x ,
\end{equation}
and
\begin{equation}\label{betaid}
    \beta = \frac{\Cdissip{H_{\rm e}}{S_{\rm e}}^Q_x}{2 \kB
    \Cdissip{H_{\rm e}}{H_{\rm e}}^Q_x} .
\end{equation}
Whereas $\beta$ can be identified uniquely, only the combination $\gamma\alpha^2$ is fixed by the comparison, not the parameters $\alpha$ and $\gamma$ separately. This is a direct consequence of a scaling property of the thermodynamic master equation (\ref{GENERICme}): any rescaling of the coupling operator $Q$ can be compensated by a corresponding rescaling of the dissipative bracket so that $Q$ does not have an absolute meaning. From eqs.~(\ref{jumpoperatorgen}) and (\ref{jump}) one realizes that the normalization of $\tilde{Q} \psi$ is proportional to $\alpha$. We can hence choose $\alpha$ such that the normalization of the state vector $\psi$ does not change on average during a jump process, that is,
\begin{equation}\label{normalization1}
    {\rm tr} ( \tilde{Q} \rho \tilde{Q}^\dag ) = 1 ,
\end{equation}
thus obtaining an absolute meaning of the jump operator $\tilde{Q}$ (and of the coupling operator $Q$). Then, the normalization of the state vector $\psi$ does not change on average during the deterministic evolution either, that is,
\begin{equation}\label{normalization2}
    {\rm tr} ( \Lambda \rho ) = 0 ,
\end{equation}
because a master equation of the form (\ref{stochme2}) preserves the trace of $\rho$ and hence, according to eq.~(\ref{unravel}), the average normalization of $\psi$. Indeed, both conditions (\ref{normalization1}) and (\ref{normalization2}) are equivalent to
\begin{equation}\label{normalization3}
    \alpha^2 \Big [ {\rm tr} ( Q \rho Q ) - \beta^2
    {\rm tr} \big(\Qcommu{Q}{H}_\rho \rho^{-1} \Qcommu{Q}{H}_\rho\big) \Big] = 1.
\end{equation}
Note that the sign of $\alpha$ is still not fixed by eq.~(\ref{normalization3}). Actually, one could even introduce a random phase in any jump without any effect on the resulting quantum master equation.

Whereas, from a theoretical point of view, eq.~(\ref{normalization3}) provides a most satisfactory condition for determining $\alpha$, for practical purposes like computer simulations it may be more convenient to work with an approximate value of $\alpha$ and to fix $\gamma$ accordingly to fulfill eq.~(\ref{gamalphid}). For example, one could choose the equilibrium value of $\alpha$  or the solution of the approximate equation $\alpha^2 {\rm tr} ( Q \rho Q ) =1$.

\subsection{Jump operator at equilibrium}
As the jump operator $\tilde{Q}$ is the key ingredient to our construction of a piecewise deterministic Markovian jump process, we next consider it in more detail for an equilibrated quantum system, that is, for \cite{hco200,hco201}
\begin{equation}\label{equilibrated}
    \rho \propto \exp \left\{ - \frac{H}{\kB T_{\rm e}} \right\}
    \quad \mbox{and} \quad
    T_{\rm e} \Cdissip{H_{\rm e}}{S_{\rm e}}^Q_x =
    \Cdissip{H_{\rm e}}{H_{\rm e}}^Q_x .
\end{equation}
By means of eq.~(\ref{Arhoident}), we obtain the following simplified expression for $\tilde{Q}$,
\begin{equation}\label{jumpoperator1}
    \tilde{Q} = \alpha \left( Q - \frac{1}{2} \Qcommu{Q_\rho}{\ln\rho} \rho^{-1} \right)
    = \frac{\alpha}{2} \, ( Q + \rho Q \rho^{-1} ) ,
\end{equation}
together with
\begin{equation}\label{Lambdaeq}
    \Lambda = \frac{\gamma}{2} \left [ 1 - \frac{\alpha^2}{4}
    \big( 3Q^2 - \rho Q^2 \rho^{-1} + Q \rho Q \rho^{-1} + \rho Q \rho^{-1} Q \big) \right] .
\end{equation}

By evaluating the matrix elements of $\tilde{Q}$ for the eigenstates $\Dket{n}$ of the Hamiltonian with eigenvalues $E_n$,
\begin{equation}\label{jumpoperator2}
    \Dbra{m} \tilde{Q} \Dket{n} = \frac{\alpha}{2} \left( 1 +
    \exp \left\{ \frac{E_n-E_m}{\kB T_{\rm e}} \right\} \right) \Dbra{m} Q \Dket{n} ,
\end{equation}
we obtain an illuminating interpretation of the jump operator $\tilde{Q}$: The coupling operator $Q$ describes the possible jumps between quantum states and the modified operator $\tilde{Q}$ introduces the proper transition probabilities by a \emph{generalized form of detailed balance} for matrix elements rather than transition probabilities. This situation is very similar to the well-known procedure in classical Monte Carlo simulations where one first defines the allowed moves and then introduces transition probabilities according to the Metropolis method in order to satisfy detailed balance (see, for example, p.~12 of \cite{Mouritsen} or p.~337 of \cite{hcobet}).

\subsection{Harmonic oscillator at equilibrium}
For example, for the one-dimensional damped harmonic oscillator with frequency $\omega$, the coupling operator $Q$ coincides with the position operator \cite{hco201} and hence describes jumps into one of the two neighboring energy states. The energy differences $E_n-E_m$ in eq.~(\ref{jumpoperator2}) are given by $\pm \hbar \omega$ and the equilibrium detailed balance factors favor jumps into the lower energy states over jumps into higher energy states. However, these jump rates are completely different from those of the Pauli master equation for the populations of the energy levels (see, for example, sec.~3.4.6.1 of \cite{BreuerPetru}) because also the deterministic influence of $\Lambda$ in eq.~(\ref{deterministic}) causes transitions between the energy eigenstates.

\section{Summary and conclusions}
We have introduced a piecewise deterministic Markovian jump process in Hilbert space with the jumps (\ref{jump}) and the deterministic evolution (\ref{deterministic}). In these equations, the jump operator $\tilde{Q}$ is defined in eq.~(\ref{jumpoperatorgen}) and the modification of the Schr\"odinger equation with Hamiltonian $H$ is given by the operator $\Lambda$ in eq.~(\ref{Lambdaid}). The remaining free parameters $\alpha$, $\beta$, and $\gamma$ are defined in eqs.~(\ref{gamalphid}), (\ref{betaid}), and (\ref{normalization3}), where actually only $\beta$ and the combination $\gamma\alpha^2$ are physically relevant. For a proper choice of $\alpha$, the normalization of the state vector is assumed to be preserved on average, both for the deterministic evolution and for the jumps.

The thermodynamic origin of our Markovian jump process is reflected in the specific form (\ref{jumpoperatorgen}) of the jump operator $\tilde{Q}$. We have realized that, by construction, the jump operator $\tilde{Q}$ incorporates a generalized principle of detailed balance. The jump operator $\tilde{Q}$ and the friction operator $\Lambda$ are connected by a fluctuation-dissipation relation.

The evolution of the second moments of the Markovian jump process is governed by the nonlinear thermodynamic quantum master equation (\ref{GENERICme}). The proposed stochastic process hence provides an unraveling of the thermodynamic master equation. Such a stochastic description of dissipative quantum systems in contact with a classical environment is very useful, both for developing stochastic simulation techniques and for gaining fundamental insights.

As a consequence of the nonlinear nature of the thermodynamic quantum master equation, the simulation of stochastic trajectories in Hilbert space according to eqs.~(\ref{jump}) and (\ref{deterministic}) requires knowledge of the evolving density matrix, that is, the second moment (\ref{unravel}) of the stochastic process. This moment can be obtained as a running ensemble average. This fundamental difference compared to linear quantum master equations does not cause any severe problems, as is well-known from the theory of nonlinear Fokker-Planck equations, which are also known as nonlinear diffusions, processes with mean-field interactions, or weakly interacting stochastic processes (see \cite{Frank}, sec.~3.3.4 of \cite{hcobook}, or secs.~3.7.2--3.7.4 of \cite{BreuerPetru}). In every time step, a single ensemble average needs to be evaluated in order to propagate all stochastic trajectories of the ensemble. Otherwise, the background and benefits of stochastic simulations are identical to the ones employed with impressive success for linear quantum master equations (see, for example, secs.~6.3 and 6.4 on photodetection and sec.~7.3 on the damped harmonic oscillator and driven two-level systems in the textbook \cite{BreuerPetru}).

From a fundamental point of view, the stochastic reformulation might lead to a deeper understanding of the evolution of dissipative quantum systems by offering a different perspective. In particular, the reformulation might offer a possibility to build a theory of two- and multi-time correlations. This would, of course, require that among the many possible unravelings of the nonlinear thermodynamic master equation there exists a physically preferable one. It has been shown in \cite{BreuerPetru95a,BreuerPetru95b} that ``the synthesis of the continuous Schr\"odinger-type evolution and the discontinuous quantum jumps of the Bohr picture'' arises naturally and directly in the description of open quantum systems, without any reference to the quantum master equation. A direct thermodynamic formulation of the piecewise deterministic Markovian jump process in Hilbert space might help in identifying a unique physically convincing stochastic process associated with a thermodynamic quantum master equation.

\acknowledgments
I wish to thank Francesco Petruccione for encouraging me to look deeper into the properties of thermodynamic quantum master equations.


\end{document}